\documentclass[aps,prd,twocolumn,amsmath,amssymb,amsfonts,nofootinbib,superscriptaddress,altaffilletter]{revtex4-2}

\usepackage{graphicx}
\usepackage{hyperref}
\hypersetup{
  bookmarksopen=true
}
\usepackage{ulem}
\usepackage{amssymb}
\usepackage{amsmath}
\usepackage{color}
\usepackage{supertabular}
\usepackage{dcolumn}
\usepackage[mathscr]{eucal}
\usepackage{mathrsfs}
\usepackage{siunitx}

\def\sci#1#2{#1\times10^{#2}}

\begin{document}

\title{First loosely coherent search for continuous gravitational wave sources with substellar companions in the Orion spur}

\author{Vladimir Dergachev}
\email{vladimir.dergachev@aei.mpg.de}
\affiliation{Max Planck Institute for Gravitational Physics (Albert Einstein Institute), Callinstrasse 38, 30167 Hannover, Germany}

\author{Maria Alessandra Papa}
\email{maria.alessandra.papa@aei.mpg.de}
\affiliation{Max Planck Institute for Gravitational Physics (Albert Einstein Institute), Callinstrasse 38, 30167 Hannover, Germany}
\affiliation{Leibniz Universit\"at Hannover, D-30167 Hannover, Germany}

\begin{abstract}
We report on the first loosely coherent search for binary systems. We searched $0.06$\,rad disk in the Orion spur, covering gravitational wave frequencies from 100 to 700\,Hz and frequency derivatives between $-10^{-11}$ to $10^{-11}$\,Hz/s. A follow-up was performed, which found no outliers. An atlas of results from the first stage of the search is made publicly available.
\end{abstract}

\maketitle

\section{Introduction}

All-sky searches for continuous gravitational waves from isolated stars over extended observation periods are computationally demanding. As optimal search methods are prohibitively expensive, sub-optimal methods are employed, requiring trade-offs between search depth, breadth, and robustness. The problem becomes harder when the emitting object rather than being isolated is in a binary system, because the orbital parameters now also need to be searched over. The result is that, with the same computational budget, an all-sky search for emission from a star in a binary system is less sensitive than an all-sky search from an isolated object over the same set of possible signal frequencies.

Previous wide parameter searches for binary systems \cite{Amicucci:2025aiz,Mirasola:2024kll,Covas:2024nzs,Covas:2022rfg, keith_review} have employed various techniques \cite{twospect1,binaryhough1,Covas:2022mqo}. Loosely coherent algorithms \cite{loosely_coherent, loosely_coherent2, loosely_coherent3} analyze patches of parameter space, gaining efficiency as the patch size increases. One would therefore expect that loosely coherent methods would be particularly suitable to tackle binary searches  because of the larger dimensionality of the waveform manifold compared to searches for isolated sources. In this paper, we present the results of the first loosely coherent search for continuous gravitational wave signals from neutron stars in binary systems with substellar companions.

The Falcon pipeline, which employs loosely coherent methods, has enabled very sensitive searches for continuous waves from isolated neutron stars with small ellipticities \cite{o3a_atlas1,o3a_atlas2,o3_atlas3}. The search presented here is the first step in extending Falcon’s capabilities to include binary systems, scaling computational performance and internal structures to accommodate the higher-dimensional parameter spaces required.

We conducted a wide-band search of a portion of the sky within the Orion Spur, restricting the modulation depth to substellar companions. This is a physically interesting region that Falcon can search with a comparatively small amount of computational resources - less than 1.2\;million CPU-thread hours, including follow-up analysis.

The follow-up analysis produced no outliers. No outliers were expected from hardware injected simulated signals because none were present in the searched sky area.

\begin{figure}[htbp]
\includegraphics[width=3.3in]{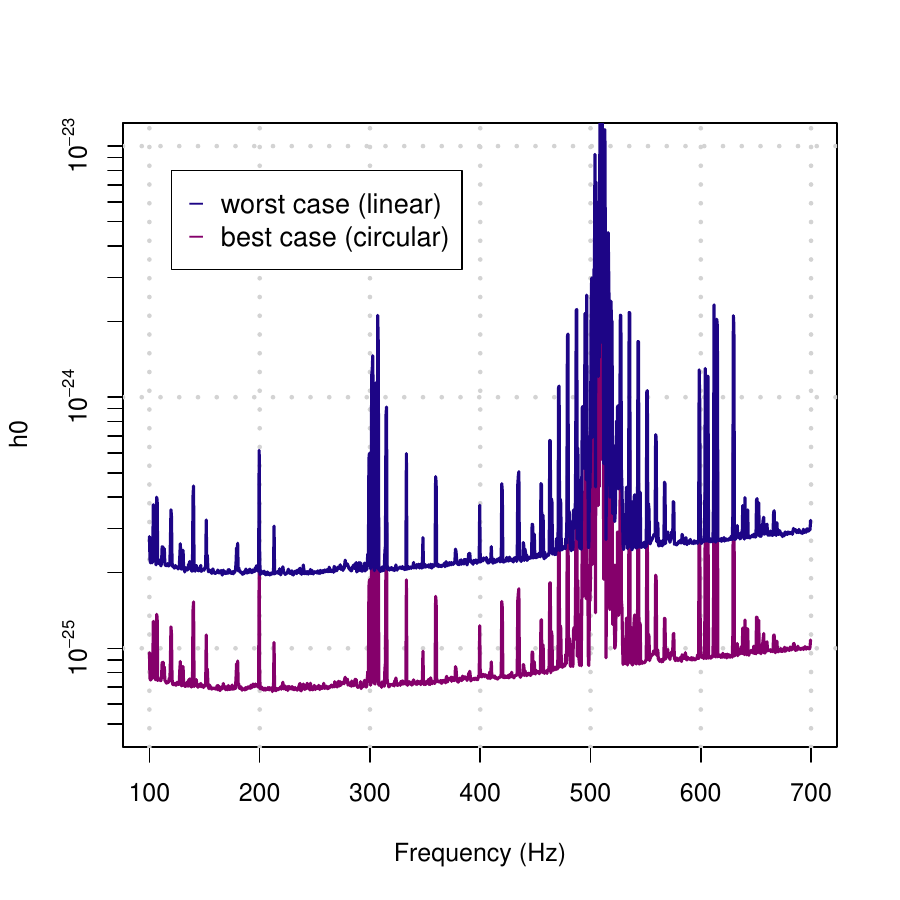}
\caption[Upper limits]{
\label{fig:amplitudeULs}
Gravitational wave intrinsic amplitude $h_0$ upper limits at 95\% confidence as a function of signal frequency. The upper curve shows worst-case upper limits, maximized over sky and polarization parameters. The lower curve shows upper limits maximized over sky for circular polarization only.
}
\end{figure}

The results of the first analysis stage were compiled into an atlas \cite{o3a_atlas1}, which we make available with this paper. A summary of the atlas is shown in Figure \ref{fig:amplitudeULs}.

\section{Search setup}
The first analysis stage and the first follow-up stage use data from the O3a LIGO public data set \cite{O3aDataSet}. The final follow-up stage uses the full public data of the O3 run \cite{O3aDataSet, O3bDataSet}. We begin with a coherence length of one hour, which is doubled in the follow-up stages.

The Orion Spur has been previously searched for continuous gravitational wave emission for isolated neutron stars \cite{orion_spur}. For this search, we select direction A as used in paper \cite{orion_spur} with right ascension of $5.2836$~rad and declination of $0.5857$~rad.

\begin{figure}[htbp]
\includegraphics[width=3.3in]{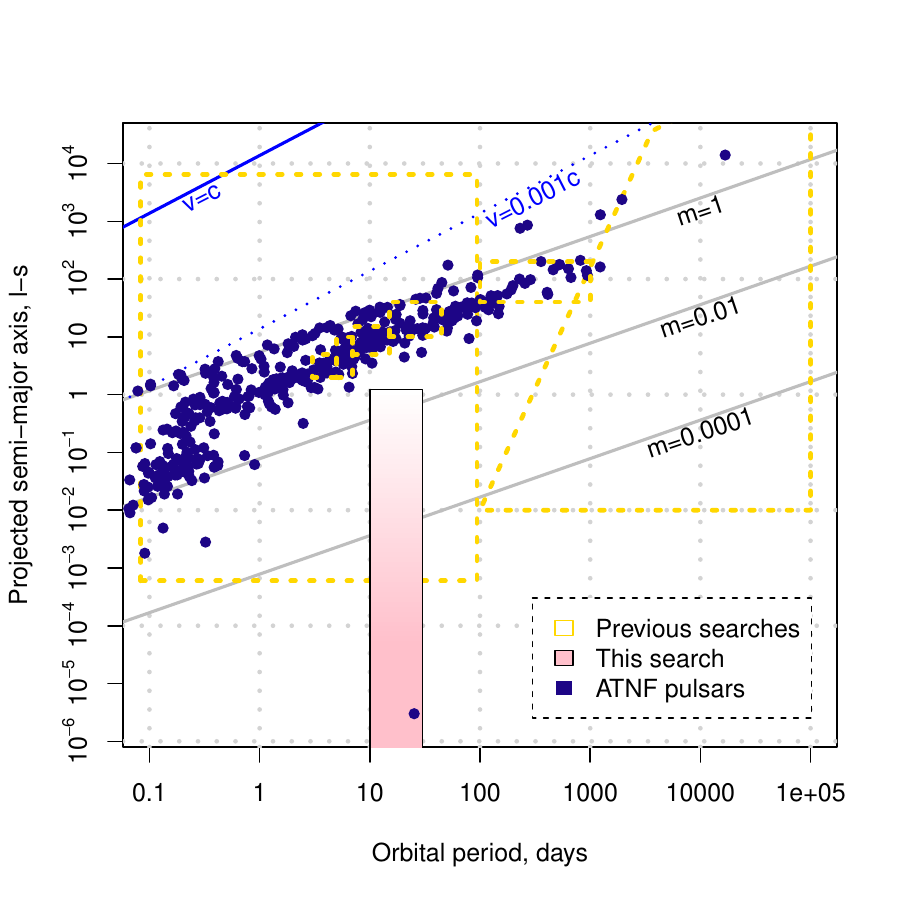}
\caption[Orbital parameter space surveyed so far]{
\label{fig:binSpace}
Orbital parameters of known pulsars (circles), regions of orbital parameter space surveyed so far by continuous wave searches (dashed yellow lines) and by this search (shaded region). The sensitivities of previous searches vary considerably, for example, the search convering the large rectangle was performed on pre-Advanced LIGO data. We also show contours of constant companion mass $m$ and the contours of companion orbital velocity $v$.
}
\end{figure}

We restrict our consideration to binary systems with circular orbits, as the orbits tend to circularize over time. We search binary periods 
\begin{equation}
10~{\textrm{days }}\leq P_{orb} \leq 30~{\textrm{days }}
\end{equation}
and projected semi-major axes
\begin{equation}
a_p \leq 1~{\textrm{l-s}} ~\left [\frac{f}{575.25~{\textrm{Hz}}}\right ],
\end{equation}
with $f$ being the signal frequency. Assuming the source of gravitational waves at 575.25\,Hz is a neutron star of $1.4$ solar masses, this covers the companions with masses from 0 to $\approx \sci{2}{-2}$ solar masses. Such binary systems are known to exist \cite{ATNF} and lie in a binary parameter space region that has not been covered by previous searches (see Figure \ref{fig:binSpace} adapted from figure 2 of \cite{Covas:2024nzs}). We use the binary signal model described in \cite{blandford_teukolsky}.

The binary parameter space was partitioned into ten equal pieces based on a single search parameter: the phase of the binary orbit at the reference time $t_0=1246070000$~GPS seconds. The atlas contains separate entries for each piece. Because the search range for  the semi-major axis includes $0$, a signal from an isolated neutron star will appear in each of the ten pieces, while a binary signal will have a peak in one of them.

The frequency and frequency derivative search ranges are
\begin{equation}
\label{eq:fRange}
100 ~{\textrm{Hz}} \leq f \leq 700 ~{\textrm{Hz}} 
\end{equation}
\begin{equation}
\label{eq:fdotRange}
-10^{-11}~[{\textrm{Hz/s}}] \leq \dot{f }\leq 10^{-11}~[{\textrm{Hz/s}}] .
\end{equation}
The upper bound of the frequency range was chosen to be high enough to include the highly contaminated frequency space around 500\,Hz as well as the cleaner spectrum above, while not being too high to avoid excessive computational resource requirements. The lower frequency range stopped at 100\,Hz to avoid the noisier low-frequency bands, which would be better analyzed with finer sky resolution.

\begin{table}[htbp]
\begin{center}
\begin{tabular}{rD{.}{.}{2}D{.}{.}{3}r}\hline
Stage & \multicolumn{1}{c}{Coherence length (hours)} & \multicolumn{1}{c}{Minimum SNR} & \multicolumn{1}{c}{Dataset} \\
\hline
\hline
1  & 1 & 9 & O3a\\
2  & 2 & 10 & O3a\\
3  & 2 & 12 & O3a+O3b \\
\hline
\end{tabular}
\end{center}
\caption{Parameters for each stage of the search.  Stage 3 refines outlier parameters by using denser sampling of the parameter space and subjects them to an additional consistency check by comparing outlier parameters from analyses of individual interferometer data. Only the first stage was used to construct the atlas, while subsequent stages were used for outlier analysis.}
\label{tab:pipeline_parameters}
\end{table}

Follow-up analysis was restricted to sky positions with signal-to-noise ratio (SNR) of at least 9 (Table \ref{tab:pipeline_parameters}).

\section{Testing}

\begin{figure}[]
\includegraphics[width=3.3in]{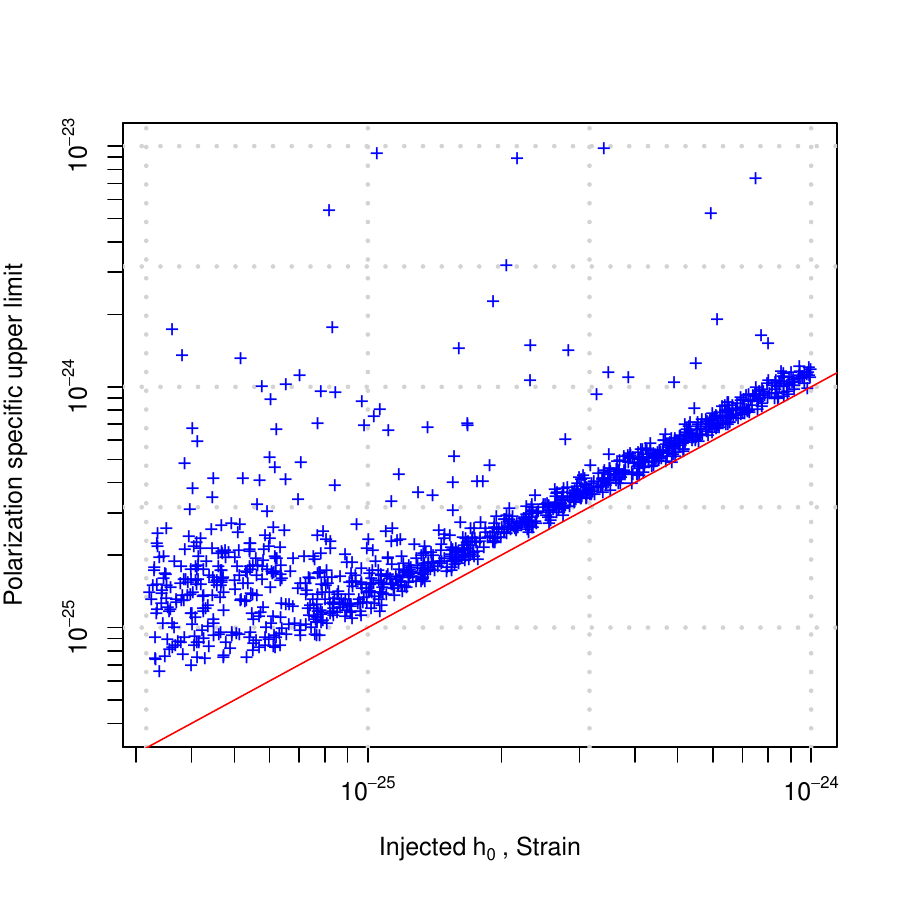}
\caption[Upper limit vs injected strain]{
\label{fig:upper_limit_recovery}
Polarization-specific upper limits obtained for test signals. The red line at the bottom is a diagonal $y=x$.
}
\end{figure}

As expected, the Falcon pipeline required modifications to adapt to the higher dimensional parameter spaces of the binary search. To validate its performance, we carried out several simulations where test signals were added to the O3 data and recovered with the new pipeline.
\begin{figure}[]
\includegraphics[width=3.3in]{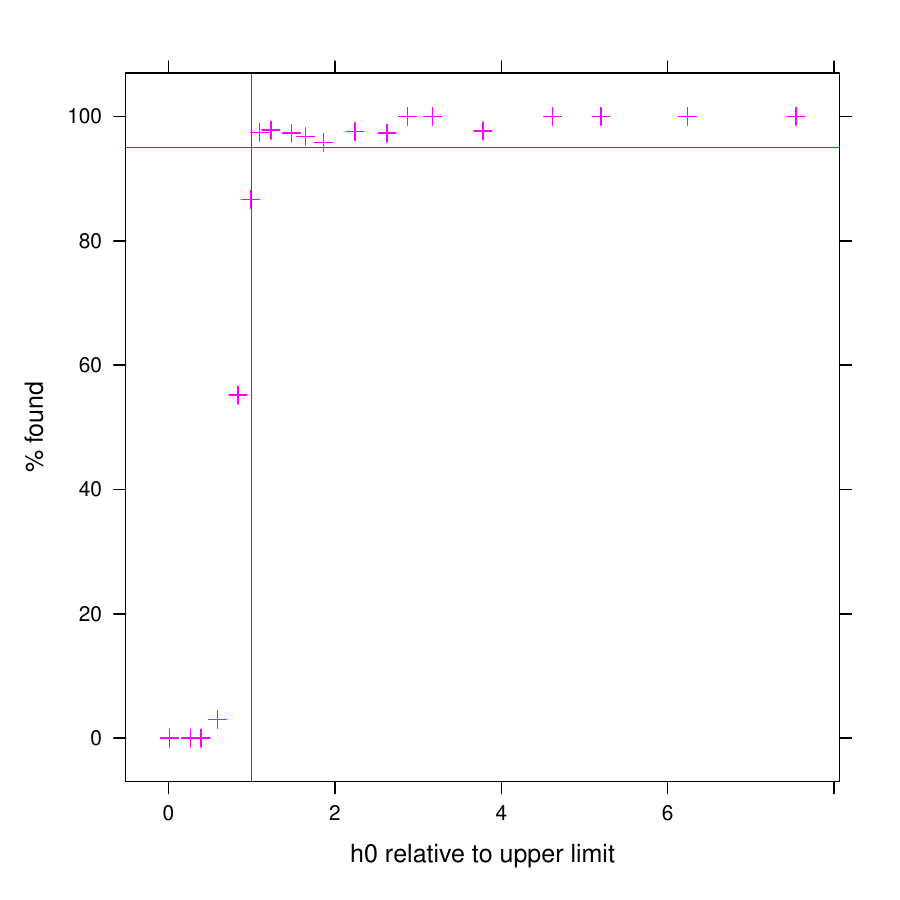}
\caption[Software injection recovery]{
\label{fig:injection_recovery}
Recovery of test signals. The X axis shows test signal strain normalized to the worst-case upper limit obtained in absence of test signals. The Y axis shows the percentage of injections recovered. The vertical green line indicates when the test signal strain is equal to the upper limit obtained without test signal. The horizontal green line shows the 95\% recovery level.
}
\end{figure}
Figure \ref{fig:upper_limit_recovery} shows the results of an extensive validation study, where we computed polarization-specific upper limits \cite{functional_upper_limits} and compared them with the strain of the added test signals. Polarisation-specific upper limits were obtained by the same procedure \cite{universal_statistics} as used in the actual analysis and contributing to the  atlas. As we can see, the upper limits exceed the test-signal amplitudes. Worst-case and circularly polarized upper limits were verified as well.

The test signals were further used to verify follow-up performance. Figure \ref{fig:injection_recovery} demonstrates that we recover 95\% of test signals above the worst-case upper limit level computed for the data without the test signals.

\section{Results}
The results obtained from our analysis have been compiled into a gravitational wave atlas \cite{data}. A summary of its data is shown on Figure \ref{fig:amplitudeULs}. In our most sensitive region, we reach worst-case strain upper limit of $\sci{2}{-25}$. This is competitive with the searches in comparable frequency regions \cite{Covas:2022rfg,LIGOScientific:2020qhb}, albeit on different orbital parameters.

\end{document}